\documentclass{llncs}
\usepackage[export]{adjustbox}

\usepackage{graphicx}
\usepackage{url}

\usepackage{listings}
\usepackage[usenames,dvipsnames]{color}

\usepackage[utf8]{inputenc}
\usepackage[T1]{fontenc}
\usepackage{color}
\usepackage{bibentry}

\usepackage{xspace}

\usepackage{algorithmic}

\usepackage{url}
\urldef{\autoseersREF}\url{FCOMP-01-0124-FEDER-020484}
\urldef{\ssaappsREF}\url{FCOMP-01-0124-FEDER-010048}
\urldef{\gslsREF}\url{POCI-01-0145-FEDER-016718}
\urldef{\crosssREF}\url{FCOMP...}
\urldef{\haslabREF}\url{FCOMP-01-0124-FEDER-022701}
\urldef{\autoseersgrant}\url{PTDC/EIA-CCO/116796/2010}
\urldef{\mygrant}\url{SFRH/BPD/73358/2010}
\urldef{\tiagosgrant}\url{SFRH/BD/30215/2006}
\urldef{\ssaappsgrant}\url{PTDC/EIA-CCO/108613/2008}
\urldef{\crosssgrant}\url{PTDC/EIA-CCO/108995/2008}
\urldef{\jorgesgrant}\url{BI2-2012_PTDC/EIA-CCO/108613/2008_UMINHO}
\urldef{\myhomepage}\url{http://www.di.uminho.pt/~jacome}
\urldef\joaosgrant\url{SFRH/ BPD/46987/2008}
\urldef\ssaappsgrant\url{PTDC/EIA-CCO/108613/2008}
\urldef\gslsgrant\url{PTDC/EEI-ESS/5341/2014}
\urldef\ssaappurl\url{http://ssaapp.di.uminho.p}
\urldef\tlturl\url{http://code.google.com/p/2lt}
\urldef\jpaulo\url{jpaulo@{di.uminho.pt,fe.up.pt}}
\urldef\jacome\url{jacome@di.uminho.pt}
\urldef\jas\url{jas@di.uminho.pt}
\urldef\jorge\url{jorgemendes@di.uminho.pt}

\usepackage[all]{xy}
\newcommand{\arLaw}[5]{%
\xymatrix{
        #1      \ar@@/^1pc/[rr]^-{#4} &
        #5 &
        #2      \ar@@/^1pc/[ll]^-{#3}
}}

\newcommand{\arLawd}[5]{%
\xymatrix{
        #1      \ar@@/^1pc/[dd]^-{#4} \\
        #5  \\
        #2      \ar@@/^1pc/[uu]^-{#3}
}}

\newcommand{\arLawo}[5]{%
\xymatrix{
        #1 \ar@@/^1pc/[ddrr]^-{#4} & & \\ 
      & #5 & \\
    & & #2 \ar@@/^1pc/[uull]^-{#3}
}}

\begin{document}

\mainmatter  

\title{Green Software Lab:\\ 
Towards an Engineering Discipline for Green Software 
\\ ~ \\ 
Project's Final Report}

\titlerunning{GSL: Green Software Laboratory ~  Project's Final Report}

%

\author{July 2021 \\ ~ \\ Rui Abreu,
  Marco Couto,
  Luís Cruz,
J{\'a}come Cunha, \\
Jo{\~a}o Paulo Fernandes,
Rui Pereira,
Alexandre Perez,
Jo{\~a}o Saraiva\inst{1,2}}
%
\authorrunning{J. Saraiva \textit{et al.}}

\institute{Department of Informatics, University of Minho\\
  \and HASLab/INESC TEC \\ Portugal\\
  \email{saraiva@di.uminho.pt}
}

\maketitle

\begin{abstract}
This report describes the research goals and results
 of the Green Software Lab (GSL) research project. This
was a project funded by Fundaç{\~ao} para a Ci\^encia e a Tecnologia (FCT) -- the Portuguese research foundation -- under reference \gslsREF,
that ran from January 2016 till July 2020.

This report includes the complete document reporting the results
achie\-ved during the project execution, which was submitted to FCT
for evaluation on July 2020. It describes the goals of the project,
and the different research tasks presenting the deliverables of each
of them. It also presents the management and result dissemination work
performed during the project's execution. The document includes also a
self assessment of the achieved results, and a complete list of
scientific publications describing the contributions of the
project. Finally, this document includes the FCT evaluation report.
\end{abstract}

\section{Aims and Work Plan}

The current widespread use of non-wired but powerful computing devices, such as,
for example, smartphones, laptops, etc., is changing the way both computer manufacturers and software engineers develop their products. In fact, computer/software
performance (ie, execution time), which was the primary goal in the last century, is
no longer the only and main concern. Energy consumption is becoming an increasing
bottleneck for both hardware and software systems. Recently, hardware manufacturers and researchers have developed techniques to reduce energy consumption mainly
focused on developing and optimizing their hardware. This is the natural approach to
reduce energy consumption since it is the hardware which literally consumes energy.
However, very much like how a driver operating a car can heavily influence its fuel
consumption, the software which operates such hardware can drastically influence its
energy consumption too~\cite{fernandes2017assisting,silvano2016antarex}!

Recent research in software engineering has defined powerful techniques to improve software developers productivity by providing, for example, advanced type and
modular systems, integrating developing environments (IDE), testing and debugging
frameworks and tools, etc. Moreover, compiler construction techniques were developed to improve the execution time of our software, namely by using partial and/or
runtime compilation, advanced garbage collectors, parallel execution, etc. All these engineering techniques and tools aim at helping software developers quickly define correct programs with optimal runtime. Unfortunately, none of these techniques nor tools
have been adapted to support greenware software development. Indeed, there is no
software engineering discipline providing techniques and tools to help software developers to analyze, understand and optimize the energy consumption of their software!
As a consequence, if a developer notices that his/her software is responsible for a large
battery drain, he/she gets no support from the language/compiler he/she is using.
In this project, we aim to study, develop, and apply methods to analyze “energy
leaks” in software source code. Thus, the focus of the project is to reason about energy
consumption at the software level. In this context, we define energy leaks as an abnormal and excessive consumption of energy by a software system. We will start the
project by adapting well known techniques for fault localization and program debugging in software source code, to locate energy-leaks in software and to relate such leaks
to the software source code. Software source code metrics and a catalog of program
smells will be adapted to the energy-aware realm.

Using these techniques, we will identify what programming practices, design pat-
terns, and other factors contribute to high energy consumption. Being able to locate
such energy leaks in the developer’s code, we will construct both a catalog of software
energy metrics and a catalog of red smells (i.e., energy inefficient smells in source code).
These techniques are the main building blocks for the next phase: providing a set of
source code refactorings and supporting tools that help developers in green decision
making and in optimizing the code. A source code refactor is a source-to-source transformation that does not change the semantic behaviour of a program. Our energy-aware
catalog of refactorings, named “green refactorings”, will be used to improve software
energy consumption. First, red smells are marked in the source code, and one or more
green refactoring will be suggested so that developers make their software greener. We
envision an Integrated Energy-aware Development Environment (IEDE) in which the
refactorings are automatically applied, or semi-automatically applied with the developer’s guidance.

To validate our techniques we will develop libraries and tools to support green de
ision making such as generic and reusable libraries implementing the catalog of energy metrics, red smells and green refactorings. To analyze and locate energy leaks,
an energy profiler, a red smell detector, and other energy monitorization tools, will be
developed. To optimize energy leaks, a framework implementing the red smells/green
refactorings will be defined as IDE plugins (Eclipse, etc.). Such a framework will localize where in the source code is a red smell, while also providing the programmer
the relevant information to show where and what is making his/her code energy inefficient, methods/alternatives, and automatically optimize the energy and refactor the
code. All of this will allow programmers to finally become energy-aware while pro-
gramming and consider this aspect of their code, and with the appropriate tools, can
finally have ways to support green decision making. Finally, we will validate our methods, tools, and techniques with real case studies with our industrial partners through
benchmarks, and professional programmers through empirical studies.

This project proposes to study and develop techniques and tools to help software
developers to analyze, to understand, and, finally, to optimize the energy consumption
of their software. The overall goal of the project is to define an engineering discipline to
support green decision making while developing software. To be more precise we wish
to answer the following Research Questions (RQ):

\begin{itemize}

\item RQ 1: Can we interpret abnormal energy consumption as software faults? And, can
we adapt advanced fault localization techniques to locate energy leaks in source code?

\item 
RQ 2: Can we develop tools to help software developers in green decision making?
Can we provide a catalog of green refactorings as IDEs plug-ins that help in developing
greener software?

\item 
RQ 3: How precise are such techniques in detecting abnormal energy consumption
in the source code of software systems? How efficient are green refactorings in optimizing energy leaks?
\end{itemize}

This document is intended to describe in detail the development of the GSL project.
In section 2 we describe the contributions and deliverables of each of the tasks in the
original plan. In Section 3 reports on management and dissemination activities. Section 4 discusses what, in the proponent’s opinion, has been achieved within GSL and
enumerates a few future research directions. Finally, Section~\ref{sec:fct} includes the project's evaluation report produced by FCT.

\section{Work Plan Execution}\label{sec:workplan}

\subsection{Task T1:  Methods for Energy Leak Localization}\label{sec:t1}

The team started the project by defining methods for locating abnormal energy consumption in the source code of a software system. First, we defined generic fault localization techniques~\cite{alexandrePHD} to locate faults in the source code of a software system. Moreover, we extended this technique   to detect energy hot-spots in the source code of a software system~\cite{Pereira20}.  This technique was named Spectrum-based Energy Leak Localization and was \textbf{awarded the silver medal} in ACM Student Research Competition (ACM SRC) at International Conference on Software Engineering (ICSE 2017) - Buenos Aires, Argentina, May 2017: Rui Pereira - a PhD students and a member of GSL - won this prestigious award~\cite{pereira2017locating}. This technique to detect energy leaks in software systems was also presented at ICSE as a poster paper~\cite{Pereira2017} and implemented as a software prototype - named SPELL - that is now available to the research community, fully defined within a prestigious journal paper~\cite{Pereira20}. 

The team also develop relevant work in analyzing the energy behavior of data structures in the overall energy
consumption of software systems. We analyze in great detail the energy consumption of the Java Collection Framework~\cite{Pereira2016} - the standard Java data structure library - and the advanced functional data structures in the Haskell programming language~\cite{lima2019jss,melfe2018sblp}. 

Together with data structures, algorithms are foundational for programmers and programming. In the project, we have also focused on studying the energy efficiency of algorithms. Common folklore abounds suggesting that a faster algorithm will always consist of a more energy-efficient one. We have confirmed that this is not always the case in an empirical way, having observed that the total energy consumed by an algorithm can be modelled as a linear combination of the energy consumed by the CPU instructions and memory accesses. Indeed, in the MSc thesis~\cite{goncalomsc}, we have studied matrix transposition algorithms, and observed that different memory access patterns and the number of activated cores have a strong influence on the energy consumption of the studied algorithms.

Software applications  have to run in a variety of mobile devices where energy consumption is becoming the bottleneck in terms of software performance. Software Product Lines (SPL) have emerged as an important software engineering discipline allowing the development of software that shares a common set of features. Thus, SPLs are particularly suitable to develop software where individual products target specific computing architectures/devices, while sharing common software features. In this task  we developed techniques to reason about energy consumption in the context of software product lines: it statically predicts the energy consumed by all products in a SPL in the worst-case scenario which we named \textit{Worst-Case Energy Consumption}.  Our results show that it is possible to accurately estimate the energy consumed by a product without actually executing and measuring it~\cite{splc17}.

We explored a well known concept in software development, technical debt, and introduced a new concept called energy debt. Energy debt is a new metric, reflecting the implied cost in terms of energy consumption over time, of choosing a flawed implementation of a software system rather than a more robust, yet possibly time consuming, approach. Similar to technical debt, if energy debt is not properly addressed, it can accumulate an energy “interest”. This interest will keep increasing as new versions of the software are released, and eventually reach a point where the interest will be higher than the initial energy debt. This work is published in~\cite{energydebt2020}, and was performed in accordance to MSc thesis~\cite{danielmsc}.

In order to measure energy consumption of software systems we purchased several equipment to monitor energy consumption, namely, a specific server supporting the Intel RAPL energy estimation framework, an ODroid board which contains energy sensors, and the external device \textit{monsoon} to monitor the energy consumed by any computer device (including mobile devices).

\vspace{0.5cm}

\noindent
\textit{Deliverables:}

\begin{itemize}

\item \textit{2 PhD thesis:} \cite{alexandrePHD}, \cite{pereiraPHD}.   

\item \textit{4 MSc thesis:} \cite{gilbertomsc}, \cite{goncalomsc}, \cite{danielmsc},  \cite{hugomsc}. 

\item \textit{14 papers:} \cite{splc17}, \cite{energydebt2020}, \cite{Lima2016}, \cite{lima2019jss}, \cite{melfe2018sblp}, \cite{Pereira2017}, \cite{pereira2017locating}, \cite{Pereira2016}, \cite{Pereira20}, \cite{Pereira2017SLE}, \cite{scp2021}, \cite{cruz2019footprint},  \cite{Cruz2019ICSME},  \cite{cruz2019attention}.

\end{itemize}

%

\subsection{Task T2: Methods for Software Energy Optimization}

In the context of this task the team developed several methods and tools to optimize energy consumption.  First, and based on the techniques for analysing energy consumption of software systems developed in the previous task (which ran in the first year of the project), we proposed a technique to optimize Java software systems by refactoring the source code so that it uses the most energy efficient Java data structures available for that particular program running on a
specific hardware architecture. Such transformation does not change the semantic of the software, but it does improve its energy consumption~\cite{Pereira2016}. We develop a similar optimization to help Haskell software  
 developers write energy efficient Haskell through a data-structure evaluation~\cite{melfe2018helping}. 

Performance is a main concern in mobile, but powerful computing devices. Google has published a set of best practices to optimize the performance of Android applications. However, these guidelines fall short to address energy consumption. As mobile software applications operate in resource-constrained environments, guidelines to build energy efficient applications are of utmost importance. In the context of this task, we performed extensive studies on whether or not a set of best performance-based practices have an impact on the energy consumed by Android applications~\cite{Couto2014,Cruz2017,acta2018,Cruz2018,Cruz2019,saner20,emse21}. The paper~\cite{Cruz2018} won the \textbf{best paper award} at CIbSE'18. Our extensive work on Android best programming practices has been invited to contribute with \textbf{two chapters to the book} \textit{Software Sustainability} to be published by Springer~\cite{book21,book21_jpf}.

Virtual keyboards are one the of most used software applications on mobile devices, and there are plenty of alternatives to choose from. We studied the energy performance of five of the most used virtual keyboards in the Android ecosystem.  We performed two empirical studies showing  that there exist relevant performance differences among the most used keyboards, and it is possible to save nearly 18\% of energy by replacing the most used keyboard in Android by the most efficient one. We also showed that is possible to save both energy and time by disabling keyboard intrinsic features and that the use of word suggestions not always compensate for energy and time~\cite{ruaMobileSoft20}. We also study the energy consumption of two popular and widely used web browsers. To properly measure the energy consumption of both browsers, we simulate the usage of various applications, which the goal to mimic typical user interactions and usage. Our preliminary results show interesting findings based on observation, such as what type of interactions generate high peaks of energy consumption, and which browser is overall the most efficient~\cite{sustainse20}.

Programming languages are everywhere in the world of computer science, and need no real introduction. They are essentially the driving factor of any hardware system. Due to this, we carefully analyzed the energy consumption of programming languages across a broad scale, to better understand what their energy efficiency is. Such studies were performed both for Android based programming languages, and all purpose programming languages, with their results present in~\cite{Lima2016,lima2019jss,Pereira2017SLE,scp2021}. The results of this work received a \textbf{best paper award} at the SBLP 2017 conference~\cite{Couto2017}, and also published at the SLE'17 conference~\cite{Pereira2017SLE}. An extended version of these two papers has been accepted  at the journal of Science of Computer Programming an is in press  now~\cite{scp2021}.


Function Memoization is widely used in the context of functional programming to speedup the execution time of programs: calls of a function are cached/memoized, and next calls to that function with the same arguments are obtained from the cache, so that the function is not call again.  We studied the impact in terms of energy consumption of (side-effect free) method memoization in several Android applications.  Our first  results show that memoization can greatly improve the energy efficiency of an Android application, considering that such application is prone to it, without threatening its normal functioning and/or efficiency~\cite{cibse19}.

In the context of this task we have now two PhD projects running: Francisco Ribeiro is studying how to adapt automated program repair to the green software realm~\cite{franciscoPHD}. José Nuno Macedo is adapting techniques, well-known in medicine, to program repair, namely (program) transplants and tissue/program growing~\cite{zeNunoPHD}. 

\vspace{0.5cm}

\noindent 
\textit{Deliverables:}

\begin{itemize}
\item \textit{4 PhD thesis:} \cite{luiscruzPHD}, \cite{marcoPHD}, 
                    \cite{franciscoPHD}, \cite{zeNunoPHD}. 
\item \textit{3 MSc projects:}  \cite{ruamsc}, \cite{mariomsc}, \cite{adrianomsc} . 

\item \textit{2 book chapters:} \cite{book21}, \cite{book21_jpf}.   

\item \textit{16 papers:}  \cite{Couto2014}, \cite{SantosSPK17}, \cite{Couto2017}, \cite{Cruz2017}, 
                  \cite{Pereira2017SLE}, \cite{melfe2018helping}, \cite{acta2018}, \cite{Cruz2018}, \cite{cruz2019improving}, \cite{cibse19}, \cite{Cruz2019}, \cite{ruaMobileSoft20}, \cite{saner20}, \cite{sustainse20}, \cite{scp2021}, \cite{emse21}.

\end{itemize}


\subsection{Task T3: Tools to Support Green Decision Making}

During the project we developed several energy-aware software prototype tools to validate the techniques developed on Task 1 and 2. These tools are available to the scientific community as open source software so that can be reused and improved and, thus contribute to the advance of the state of the art on green software.
In this task we integrated all these tools in a  (open source) repository and are also available from our webpage.  

The software tools we developed in the GSL projectare:

\begin{itemize}

\item  \textit{SPELL}: SPectrum-based Energy Leak Localization~\cite{Pereira20}.

\item \textit{Leafactor}: Automatic refactoring toolset for energy efficiency of Android~\cite{Cruz2017leafactor}.

\item \textit{JStanley}: Placing a green thumb on Java collections~\cite{Pereira2018}.

\item \textit{EMaaS}: Energy Measurements as a Service for Mobile Applications~\cite{cruz2019emaas}.

\item \textit{GreenHub}: A collaborative approach to power consumption analysis of Android devices~\cite{Matalonga2019} ({\bf MSR'19 Data Showcase Special Mention Award}).

\item \textit{AnaDroid}: a tool to monitor and analyse energy consumption of Android applications.

\item \textit{E-Debitum}: Managing Software Energy Debt~\cite{e-debitum}.

\end{itemize}

We also developed other software artifacts, namely repositories, and ?? which are also available 
to the scientific community: 

\begin{itemize}

\item \textit{Energy Efficiency in Programming Languages}: repository and benchmarking environment to measure the energy efficiency of 27 popular programming languages

\item \textit{GreenSource}: a repository of Android applications~\cite{Rua2019MSR}.

\item \textit{Collections Energy Benchmark}:  A benchmark to measure Java collection methods

\item \textit{Collections Energy Tables}:  Interactive energy data tables for the Java Collection Framework 
\end{itemize}

All the tools developed in the project are available from the webpage
of the project (see task 5)

\vspace{0.5cm}

\noindent
\textit{Deliverables:}

\begin{itemize}

\item 6 papers: \cite{Cruz2017leafactor}, ~\cite{Pereira2018}, \cite{e-debitum}, \cite{Rua2019MSR}, \cite{Matalonga2019}, \cite{cruz2019emaas}.

\item \textit{7 software tools:} Leafactor, jStanley, SPELL,  EMaaS, GreenHub, AnaDroid, E-Debitum.

\end{itemize}

\subsection{Task T4: GreenSoftwareLab: Validation}

In order to validate the methods and tools we developed in this project, we are conducting several empirical studies with real software developers.

We have conducted a large study with 20 software developers to validate the Spectrum-based Energy Leak Localization (SPELL) technique we developed on Task 1.  The results achieved confirm that our
technique indeed help programmers locating abnormal energy consumption in software systems~\cite{pereiraPHD,Pereira20}.

In the context of Android - the largest mobile ecosystem - several works have focused on documenting energy-greedy programming patterns and on finding better alternatives for
them. In fact, identifying and refactoring such code patterns to
improve energy consumption has already presented promising
research results. These results, however, have essentially been validated by testing code patterns individually and often in a small set of applications (sometimes only in one).
To validate those results in a large-scale repository, we considered  11 energy-greedy code patterns
obtained from the literature, and considered a repository of 600+ Android applications to understand the frequency of occurrence of such patterns. Within the 200+ applications where the patterns were detected, we studied the impact that replacing them, individually and combined, by their documented alternatives
has on the energy consumption. Moreover, as we consider all the possible combinations of the individual patterns, this resulted in 400+ refactored applications under analysis.
To perform our study, we developed an extensible, fully automated framework called \textit{Chimera}, which is able to detect and refactor the patterns. Each pattern is considered individually
and is also combined with all the other patterns. \textit{Chimera} also
measures the energy consumed by an application in different
simulated usage scenarios, before and after refactoring. This work was published at the SANER conference~\cite{saner20} and is also part of Marco Couto's PhD thesis~\cite{marcoPHD}.

In order to provide a suitable and uniform setting for evaluating software
analysis and optimization techniques in the Android ecosystem, we developed the \textit{GreenSource} infrastructure: a large body of open source Android applications tailored for energy analysis and optimization. \textit{GreenSource} consists of three main components: (1) a large collection of open source, executable
Android applications, (2) a benchmarking framework, called \textit{AnaDroid}, to test such applications under different usage scenarios and collect structural and energy-related metrics, and
(3) a large scale repository of metrics obtained from executing the applications using AnaDroid. This work was started in the context of Rui Rua's MSc thesis~\cite{ruamsc}, under a GSL research grant, and it was  published at the MSR conference~\cite{Rua2019MSR}. Rui is now a developing his PhD projet in this area supported by a FCT grant~\cite{ruaPHD}.

The greenhub initiative follows a novel initiative to analyse the energy consumption of android applications: it uses a collaborative approach where android users are invited to install the greenhub Android app and this app monitors the energy consumption of the devices in a normal environment. The greenhub app stores all
measurements in the greenhub database, which is then analysed using data mining techniques to reason about which API, hardware, settings, etc may influence the consumption of an android device. This
initiative already attracted thousands of collaborators and the media attention (see Dissemination Section).

\vspace{0.5cm}

\noindent
\textit{Deliverables:}

\begin{itemize}

\item \textit{1 PhD thesis:} \cite{ruaPHD}.    

\item \textit{1 MSc thesis:} \cite{ruamsc}.     

\item \textit{2 software tools:} \textit{Chimera}, \textit{GreenSource}~\cite{Rua2019MSR}.

\end{itemize}

\subsection{Task T5: Project Management}

In the context of this task, we developed an up-to-date webpage of the
project, where all publications, software tools, and project
activities are available to the scientific community. This webpage is
available at:

\medskip

\begin{center}
\url{http://greenlab.di.uminho.pt/fct/}
\end{center}

\medskip

As stated in the proposal, we organize an annual GSL workshop, single day event.

\begin{itemize}
    
\item  \textit{First GSL Workshop:} The first workshop was organized by João Paulo Fernandes at Universidade da
Beira Interior in Covilhã on 20$^{th}$ January of 2017.  The first GSL workshop was structured in two sessions: the morning session included presentations reporting results already achieved in the project. In the afternoon we had a brainstorming session, discussing open problems, how to organize future work, and the collaboration between the four sites of the project.
All members of the project attended this single workshop.

\item \textit{Second GSL Workshop:} The second workshop was organized by João Saraiva at Universidade do Minho
in Braga on 29$^{th}$ October of 2018. This workshop coincide with the PhD defense of the team member Rui Pereir. Thus, the program of the workshop started with a morning session consisting of Rui's PhD defense. The project's consultant Prof. Joost Visser, Radboud University, The Netherlands, was the main opponent in the PhD defense, and started the afternoon session with an invited talk with title \textit{"Green Software Research by SIG"}. After this talk, members of team gave talks where they presented their ongoing results. The workshop finished with a brainstorming session chaired by Dr. Rui Pereira. 

\item \textit{Third GSL Workshop:} The fourth workshop was organized by João Paulo Fernandes\footnote{During the execution of the project, João Paulo Fernandes moved from Universidade da Beira Interior to Universidade de Coimbra} at Universidade de Coimbra on 10$^{th}$ May of 2019. This workshop started with an invited talk by Prof. Zoltan Porkolab, Eotvos Lorand University, Budapest, Hungary, with title \textit{"Save the Earth, program in C++!"}, followed by  sessions with presentations reporting results achieved in the project.  The members of the project and several staff members of the Universidade de Coimbra attended the workshop.

\end{itemize}

In order to advertise the results achieved in the project we published an article in the Science Impact magazin, which include an interview with the PI~\cite{impact18}. A brochure of the project was produced by this magazin which we are using now to distribute in conferences attended by GSL members. In collaboration with the consultant Prof. Patricia Lago, Vrije University, Amsterdam, The Netherlands, and her research group,  the PI organized (as co-chair) the GREENS 2018 workshop, co-located with the ICSE conference~\cite{greens2018}.

\vspace{0.5cm}

\textit{Deliverables:}

\begin{itemize}

\item \textit{2 papers:} \cite{impact18}, \cite{greens2018}.

\end{itemize}

\section{Management and Dissemination}\label{sec:md}

\subsection{Management}

The management structure adopted for the GSL project follows its 
research proposal. The project was managed by the PI, who coordinated
and monitored the progress of the overall project. The PI was
responsible for organizing a weekly team meeting, that were scheduled
for Friday afternoon. In each of the meetings, with some rare
exceptions, one team member was asked to prepare 3 to 4 slides, so
that he could present his work in a more organized way. The PI was
responsible for scheduling such short/informal talks and to motivate
team members to participate.  These weekly meetings were attended by
most of the team members, and were a very important moment to discuss
new ideas and to share research results.

The PI was also responsible to coordinate the collaboration and visits of our consultants. Prof. Joost Visser and Prof. Zoltán Porkolab attended the first and third workshop of the project, respectively, where they gave keynote talks. Prof. Joost Visser was also  a member of the PhD committee of Rui Pereira PhD thesis. We also collaborated with Prof. Porkolab in the co-supervision of Mário Santos MSc thesis~\cite{mariomsc}, which was developed at Ericsson in Budapest. This resulted in co-authored research papers~\cite{SantosSPK17}. The PI visited the group of Prof. Patricia Lago at Vrij Univeristy in Amsterdam on December 15, 2017, where we decide to co-organize the GREENS workshop, a co-located event of the ICSE conference. Unfortunately, Prof. Patricia Lago had to cancel her participation at the second GSL workshop. Nevertheless, we were in close contact and ongoing collaboration throughout the project. 

To manage all the research results achieved, the software prototypes produced and to disseminate our results, we used a collaborative (wiki) site. This site includes several subpages, namely:

\begin{itemize}
\item \textit{Publications:} where all scientific papers and MSc and PhD thesis are available.

\item \textit{Software:} where all software prototypes are available for downloading

\item \textit{Activities:} where all activities of the project are described. For example, the visits of our consultants, the participation in conferences by team members, etc

\item \textit{Workshops:} we organized two SSaaPP workshops during the
  execution of the project. The program of those workshops are
  available here.

\item \textit{Empirical Studies:} the empirical studies we conducted
  produced several material that can be useful for the scientific
  community and that we refer in resulting scientific
  publications. This material is available in this sub-page.

\end{itemize}

\subsection{Dissemination}

The project scientific results were disseminated by 37 publications, and we have 3 papers under reviewing. Moreover, some of our scientific results were published at very prestigious journals (JSS, ESE, SCP, TSE) and events (ICSE, ASE, ICSME, SANER, SPLC, MSR, MobileSoft). The project let to 7 PhD thesis, 4 awarded and 3 running, and to 8 MSc thesis (7 awarded and 1 running). In two cases the work of grant holders evolved into PhD projects that will be concluded after the project lifetime. Such are the cases of grant holders Francisco Ribeiro and Rui Rua. Franciso Ribeiro is now  doing a PhD thesis on automated program repair, that will be applied to perform automated energy-aware  program repair. Rui Rua is continuing his PhD project the work on the large scale analysis of energy consumption in evolving Android applications. 

Furthermore, our research attracted the attention of the media, and we have been invited to participate in different science-oriented radio programs. This includes participation in \textit{Antena 1}'s \textit{"90 segundos de ciência"} with \textit{GreenHub: Este projeto quer ajudá-lo a poupar a bateria do seu smartphone}, and in \textit{Antena 1}'s \textit{"Ponto de Partida"} within \textit{As baterias do futuro}, both in 2018.

The project produced several software prototypes. Such prototypes are available from the project webpage. To advertise our tools, we produced tool demo videos that are available in the Internet (youtube) and accessible from the project's webpage.

Work of ours has been shown to be impactful to a wide community of researchers and practitioners; to this extent, we have included some pointers to (extensive) threads initiated upon the publication of our previous paper, showcasing the communities' (academic and industrial practitioners) openness, excitement, and interest in such findings~\cite{jax,newstack,insights,hackernews,rustNews,progCourse,twitter1,twitter2,twitter3,reddit1,reddit2}. 

Moreover, the project promoted three GSL workshops (see workshop programs in the projects webpage) where consultants gave keynote talks and team members gave talks about their work in the project. These GSL workshops were open to the research community, and in all instances of the workshop we had participants external to the project who did attend these events.

Team members were also invited to present the project's research
results at international events. The PI gave a keynote talk
on \textit{"Energyware analysis"} at the \textit{7$^{th}$ Workshop on
Software Quality Analysis, Monitoring, Improvement, and Applications}
(SQAMIA), Novi Sad, Serbia, 2018~\cite{sqamia18}.  In 2019 the PI,
João Paulo Fernandes and Rui Pereira were invited to give two
tutorials on Energy-Aware Software at the \textit{Central European PhD
School on Functional Programming} (CEFP), Budapest,
Hungary~\cite{cefp19}, \cite{cefp19edu}. Springer Verlag already
agreed to publish the tutorial papers in the final proceedings in the
LNCS Tutorial series in 2021.  Also in 2019, João Paulo Fernandes was
invited to give a keynote talk on \textit{"Ranking Programming
Languages for their Energy Efficiency"} at the \textit{8$^{th}$
Symposium on Languages, Applications and Technologies} (SLATE
2019). In 2020, the PI was an invited to be a member of the
panel \textit{Software Energy Efficiency: When Academia Meets
Industry} at the \textit{Eleventh international Green and sustainable
computing} (IGSC'20) conference (October 2020).

Team members had also important roles as organizers or program committee members of international events devoted to green software, namely:

\begin{itemize}

\item the PI and Rui Pereira serve in the program committee of the \textit{1st International Workshop on Sustainable Software Engineering} (SustainSE 2020), co-located with ASE, Melbourne, Australia, September 2020.

 \item Rui Abreu served in the program committee and Luís Cruz was the publicity co-chair of \textit{Seventh IEEE/ACM International Conference on Mobile Software Engineering and Systems}, Seoul, South Korea, May 2020.

\item The PI served in  the program committee of the \textit{17$^{th}$ International Conference on the European Energy Market} (EEM 2020), Stockholm, Sweden, September 2020.

\item João Paulo Fernandes served in the program committee of \textit{Computing for Sustainability}, a national conference co-located with INFORUM, Guimarães, Portugal, September, 2019.

\item João Paulo Fernandes served in the program committee of \textit{8th International Workshop on Requirements Engineering for Sustainable Systems} (RE4SuSy 2019), Jeju Island, South Korea, September, 2019.

\item The PI served in  the program committee of the \textit{16$^{th}$ International Conference on the European Energy Market} (EEM 2019), Ljubljana, Slovenia, September
  2019.
  
\item Luís Cruz served as Tool Demo and Mobile Applications Co-chair at the 
\textit{Sixth IEEE/ACM International Conference on Mobile Software Engineering and Systems}, Montréal, Canada, May 2019.
 
\item The PI served as Organization and Program co-chair at  the \textit{Sixth International Workshop on Green And Sustainable Software} (GREENS 2018), co-located with ICSE, Stockholm, Sweden, May 2018. João Paulo Fernandes integrated its program committee.

\item Luís Cruz served as publicity and media co-chair at  \textit{Fifth IEEE/ACM International Conference on Mobile Software Engineering and Systems}, Gothenburg, Sweden, May 2018.

\end{itemize}

\vspace{0.5cm}

\noindent
\textit{Deliverables:}

\begin{itemize}

\item \textit{3 papers:} \cite{sqamia18}, \cite{cefp19}, \cite{cefp19edu}.

\item \textit{4 keynote talks.}

\end{itemize}

\section{Self Assessment}

Globally the GSL project was very successful in meeting the proposed
research aims and following the original work plan with almost no
deviations. All scientific results and deliverables are listed above
in Sections~\ref{sec:workplan} and \ref{sec:md}.  They have far exceeded the expected indicators.

In particular, it must be mentioned the project contribution with
respect to

\begin{itemize}

\item Defining a novel technique to locate energy leaks in the source code of a software system. 

\item Defining performance-based guidelines for energy efficient mobile software applications.

\item Defining the concept of energy-debt that adapt the metaphor of technical debt to the green software realm. 

\item Defining a ranking of energy consumption for programming languages, whose research paper have more than 70 citations (according to google scholar). 

\item Defining a ranking of data structures for the Java and Haskell programming languages.

\item Defining a static analysis technique to green software reasoning in software product lines.

\item Performing empirical studies that show that our techniques do help software developers in 
in green software development, and also do optimize exiting software systems.

\item Providing several software prototypes, available as open source
  software to the scientific community, that offer energy profilers,  energyaware refactorers, and energy-debt calculators.

\end{itemize}

The novel research developed in the context of the GSL  won the following awards:

\begin{itemize}
    \item ACM Student Research Competition (ACM-SRC) at ICSE'17: Silver medal.
    \item Best paper award at SBLP'17.
    \item Best paper award at CIbSE'18.
    \item Data Showcase Special Mention Award at MSR'19.
\end{itemize}

Members of GSL  were invited to present their research results in the following international events:

\begin{itemize}
    \item Keynote Talk: \textit{"Energyaware Software"} at the workshop on Software Quality Analysis,
  Monitoring, Improvement, and Applications, SQAMIA 2019. 
    \item Tutorial Talks: \textit{"Paint your Programs Green - On the Energy Efficiency of Data Structure Implementations"} and \textit{"Green Software in an Engineering Course"} at the Central European PhD School on Functional Programming, CEFP 2019
    \item Keynote Talk: \textit{"Ranking Programming Languages for their Energy Efficiency"} at the Symposium on Languages, Applications and Technologies, SLATE 2019
    \item Panelist: \textit{Software Energy Efficiency: When Academia Meets Industry} at the \textit{Eleventh international Green and sustainable computing}, IGSC 2020). 
\end{itemize}

The work, results, and international cooperation fostered by the project planted the seeds for the involvement of the team members in new follow-up projects:

\begin{itemize}
    \item Sustainable: Promoting Sustainability as a Fundamental Driver in Software Development Training and Education, Erasmus+ Strategic Partnership (project No. 2020-1-PT01-KA203-078646), being coordinated by a GSL team member
    
    \item Cerciras: Connecting Education and Research Communities for an Innovative Resource Aware Society, COST Action CA19135, action started in 2020 and integrated by many of the team members
\end{itemize}

\vspace{.5cm}

\noindent
In terms of deliverables the project produced:

\begin{itemize}
\item 7 PhD thesis (4 awarded and 3 running);    
\item 8 MSc thesis (7 awarded and 1 running);    
\item 2 PhD School tutorials at CEFP 2019 (by invitation) with formal publication (to appear in 2021);
\item 2 book chapter (to appear);
\item 38 research papers;
\item 4 keynote talks;
\item 9 software tools; 
\item 3 GSL workshops;
\item 1 International Workshop.
\end{itemize}

All these results were obtained during the execution of the project. This includes the papers published in 2020 which report results from research developed in 2019. Actually, most of those papers where submitted in 2019.

\section{FCT Evaluation} \label{sec:fct}

The GreenSoftwareLab project was evaluated on March 2021 by the \textit{Painel de Avaliação Final - Ciências da Computação e da Informação e Informática} from FCT - the Portuguese national funding agency for science, research and technology,  which produced the following evaluation report:

\begin{quote}
"The main objectives of the GreenSoftwareLab project, which were to study and provide new methods and techniques for analysing energy leaks at the software level, as well as tools to support the development of more energy-efficient code, were fully achieved.

Project results were of high quality and resulted in several publications in good or very good quality venues (clearly exceeding, in quantity, the expectations). The research developed in the project also won a few awards, attracted the attention of media, and already received a considerable number of citations.

Several Master and PhD students defended their theses (also exceeding the expectations), addressing topics fully aligned with the project objectives.

Several tools were developed and are available through the project web page.

Overall, this project should be considered very successful, particularly considering that the requested budget was relatively low."
\end{quote}

\bibliographystyle{plain}
\bibliography{my}

\begin{thebibliography}{10}

\bibitem{reddit2}
{Reddit Examples}.
\newblock
  \url{https://www.reddit.com/search/?q=energy%20efficiency%20across%20programming%20languages}.
\newblock [Online; accessed 28-January-2021].

\bibitem{reddit1}
{Reddit Thread}.
\newblock
  \url{https://www.reddit.com/r/programming/comments/ffu7rv/will_c_ever_be_beaten_this_paper_presents_a_study/}.
\newblock [Online; accessed 28-January-2021].

\bibitem{twitter1}
{Twitter Examples 1}.
\newblock
  \url{https://twitter.com/search?q=energy%20efficiency%20across%20programming%20languages&src=typed_query&f=live}.
\newblock [Online; accessed 28-January-2021].

\bibitem{twitter2}
{Twitter Examples 2}.
\newblock
  \url{https://twitter.com/search?q=sites.google.com%2Fview%2Fenergy-efficiency-languages&src=typed_query&f=live}.
\newblock [Online; accessed 28-January-2021].

\bibitem{twitter3}
{Twitter Examples 3}.
\newblock
  \url{https://twitter.com/search?q=greenlab.di.uminho.pt&src=typed_query&f=live}.
\newblock [Online; accessed 28-January-2021].

\bibitem{Couto2014}
M.~Couto, Carção T., J.~Cunha, J.~P. Fernandes, and J.~Saraiva.
\newblock Detecting {A}nomalous {E}nergy {C}onsumption in {A}ndroid
  {A}pplications.
\newblock In Fernando~Magno Quintão~Pereira, editor, {\em Programming
  Languages}, volume 8771 of {\em LNCS}, pages 77--91. Springer Int.
  Publishing, 2014.

\bibitem{marcoPHD}
Marco Couto.
\newblock {\em Supporting Software Developers in Making Energy Saving
  Decisions}.
\newblock PhD thesis, Universidade do Minho, 2020.

\bibitem{splc17}
Marco Couto, Paulo Borba, J\'{a}come Cunha, Jo{\~a}o~Paulo Fernandes, Rui
  Pereira, and Jo{\~a}o Saraiva.
\newblock Products go green: Worst-case energy consumption in software product
  lines.
\newblock In {\em Proceedings of the 21st International Systems and Software
  Product Line Conference - Volume A}, SPLC '17, page 84–93, New York, NY,
  USA, 2017. Association for Computing Machinery.

\bibitem{energydebt2020}
Marco Couto, Daniel Maia, Jo{\~a}o Saraiva, and Rui Pereira.
\newblock On energy debt: Managing consumption on evolving software.
\newblock In {\em Proceedings of the 3rd International Conference on Technical
  Debt}, TechDebt '20, page 62–66, New York, NY, USA, 2020. Association for
  Computing Machinery.

\bibitem{Couto2017}
Marco Couto, Rui Pereira, Francisco Ribeiro, Rui Rua, and Jo{\~a}o Saraiva.
\newblock Towards a {G}reen {R}anking for {P}rogramming {L}anguages.
\newblock In {\em Proceedings of the 21st Brazilian Symposium on Programming
  Languages}, SBLP 2017, pages 7:1--7:8. ACM, 2017.
\newblock (Best {paper} {A}ward).

\bibitem{saner20}
Marco Couto, Jo{\~a}o Saraiva, and Jo{\~a}o~Paulo Fernandes.
\newblock Energy refactorings for android in the large and in the wild.
\newblock In {\em 2020 IEEE 27th International Conference on Software Analysis,
  Evolution and Reengineering (SANER)}, pages 217--228, 2020.

\bibitem{Cruz2019ICSME}
L.~{Cruz}, R.~{Abreu}, J.~{Grundy}, L.~{Li}, and X.~{Xia}.
\newblock Do energy-oriented changes hinder maintainability?
\newblock In {\em 2019 IEEE International Conference on Software Maintenance
  and Evolution (ICSME)}, pages 29--40, 2019.

\bibitem{cruz2019footprint}
Lu\'{i}s {Cruz} and R.~{Abreu}.
\newblock On the energy footprint of mobile testing frameworks.
\newblock {\em IEEE Transactions on Software Engineering}, pages 1--1, 2019.

\bibitem{Cruz2017}
Lu\'{i}s Cruz and Rui Abreu.
\newblock Performance-based {G}uidelines for {E}nergy {E}fficient {M}obile
  {A}pplications.
\newblock In {\em Proceedings of the 4th International Conference on Mobile
  Software Engineering and Systems}, MOBILESoft '17, pages 46--57. IEEE Press,
  2017.

\bibitem{Cruz2018}
Luis Cruz and Rui Abreu.
\newblock Using automatic refactoring to improve energy efficiency of android
  apps.
\newblock In {\em Proceedings of the {XXI} Iberoamerican Conference on Software
  Engineering, Bogota, Colombia, April 23-27, 2018}, pages 163--176, 2018.

\bibitem{Cruz2019}
Lu\'{i}s Cruz and Rui Abreu.
\newblock Catalog of energy patterns for mobile applications.
\newblock {\em Empirical Software Engineering}, 24(4):2209--2235, Aug 2019.

\bibitem{cruz2019emaas}
Luis Cruz and Rui Abreu.
\newblock Emaas: Energy measurements as a service for mobile applications.
\newblock In {\em 2019 IEEE/ACM 41st International Conference on Software
  Engineering: New Ideas and Emerging Results (ICSE-NIER)}, pages 101--104.
  IEEE, 2019.

\bibitem{cruz2019improving}
Luis Cruz and Rui Abreu.
\newblock Improving energy efficiency through automatic refactoring.
\newblock {\em J. Softw. Eng. Res. Dev.}, 7:2, 2019.

\bibitem{cruz2019attention}
Luis Cruz, Rui Abreu, and David Lo.
\newblock To the attention of mobile software developers: guess what, test your
  app!
\newblock {\em Empirical Software Engineering}, 24(4):2438--2468, 2019.

\bibitem{Cruz2017leafactor}
Lu\'{i}s Cruz, Rui Abreu, and Jean-Noël Rouvignac.
\newblock Leafactor: {I}mproving {E}nergy {E}fficiency of {A}ndroid {A}pps via
  {A}utomatic {R}efactoring.
\newblock In {\em IEEE/ACM International Conference on Mobile Software
  Engineering and Systems, MobileSoft 2017}, MOBILESoft '17, pages 205--206,
  2017.

\bibitem{luiscruzPHD}
Luís Cruz.
\newblock {\em Tools and Techniques for Energy-Efficient Mobile Application
  Development}.
\newblock PhD thesis, Universidade do Porto, 2019.

\bibitem{sustainse20}
J.~{de Macedo}, J.~{Aloísio}, N.~{Gonçalves}, R.~{Pereira}, and J.~{Saraiva}.
\newblock Energy wars - chrome vs. firefox: Which browser is more energy
  efficient?
\newblock In {\em 2020 35th IEEE/ACM International Conference on Automated
  Software Engineering Workshops (ASEW)}, pages 159--165, 2020.

\bibitem{book21}
Daniel Feitosa, Lu{\'i}s Cruz, Rui Abreu, Jo{\~a}o~Paulo Fernandes, Marco
  Couto, and Jo{\~a}o Saraiva.
\newblock Patterns and energy consumption: Design, implementation, studies and
  stories.
\newblock In Mario~Piattini Coral~Calero, Marian~Moraga, editor, {\em Software
  Sustainability}. Springer, (to appear).

\bibitem{fernandes2017assisting}
Benito Fernandes, Gustavo Pinto, and Fernando Castor.
\newblock Assisting non-specialist developers to build energy-efficient
  software.
\newblock In {\em 2017 IEEE/ACM 39th International Conference on Software
  Engineering Companion (ICSE-C)}, pages 158--160. IEEE, 2017.

\bibitem{cefp19}
Jo{\~a}o~Paulo Fernandes, Rui Pereira, and Jo{\~a}o Saraiva.
\newblock Paint your programs green - on the energy efficiency of data
  struc-ture implementations.
\newblock In {\em Central European Functional Programming School: 8th Summer
  School, CEFP 2013, Cluj-Napoca, Romania, June 17-21, 2019, Revised Selected
  Papers}. Springer International Publishing, (to appear).

\bibitem{insights}
Insights.
\newblock {Programming Language Efficiency Ranked: An Old Standby Wins}.
\newblock
  \url{https://insights.dice.com/2018/06/01/programming-language-efficiency-ranked/}.
\newblock [Online; accessed 28-January-2021].

\bibitem{jax}
Jaxenter.
\newblock {Java is one of the most energy-efficient languages, Python among
  least energy efficient}.
\newblock
  \url{https://jaxenter.com/energy-efficient-programming-languages-137264.html}.
\newblock [Online; accessed 28-January-2021].

\bibitem{Lima2016}
L.~G. Lima, F.~Soares-Neto, P.~Lieuthier, F.~Castor, G.~Melfe, and J.~P.
  Fernandes.
\newblock Haskell in {G}reen {L}and: {A}nalyzing the {E}nergy {B}ehavior of a
  {P}urely {F}unctional {L}anguage.
\newblock In {\em 2016 IEEE 23rd Int. Conf. on Software Analysis, Evolution,
  and Reengineering (SANER)}, volume~1, pages 517--528, March 2016.

\bibitem{lima2019jss}
Luís~Gabriel Lima, Francisco Soares-Neto, Paulo Lieuthier, Fernando Castor,
  Gilberto Melfe, and João~Paulo Fernandes.
\newblock {On Haskell and energy efficiency}.
\newblock {\em Journal of Systems and Software}, 149:554 -- 580, 2019.

\bibitem{goncalomsc}
Gonçalo Lopes.
\newblock A study on the energy efficiency of matrix transposition algorithms.
\newblock Master's thesis, Universidade de Coimbra, 2019.

\bibitem{zeNunoPHD}
José~Nuno Macedo.
\newblock {\em Medicus-Aware Software: Self-healing Software}.
\newblock PhD thesis, Universidade do Minho, ongoing: started 2018.

\bibitem{danielmsc}
Daniel Maia.
\newblock Energy debt: Applying technical debt to energy consumption.
\newblock Master's thesis, Universidade do Minho, 2020.

\bibitem{e-debitum}
Daniel Maia, Marco Couto, Rui Pereira, and Jo{\~a}o Saraiva.
\newblock {E-Debitum}: Managing software energy debt.
\newblock In {\em 1st International Workshop on on Sustainable Software
  Engineering (SUSTAIN-SE)}. (to appear), 2020.

\bibitem{greens2018}
Ivano Malavolta, Rick Kazman, and Jo{\~a}o Saraiva, editors.
\newblock {\em Proceedings of the 6th International Workshop on Green and
  Sustainable Software, GREENS@ICSE 2018, Gothenburg, Sweden, May 27, 2018}.
  {ACM}, 2018.

\bibitem{hugomsc}
Hugo Matalonga.
\newblock Extending smartphones' battery life.
\newblock Master's thesis, Universidade do Minho, ongoing: started 2019.

\bibitem{Matalonga2019}
Hugo Matalonga, Bruno Cabral, Fernando Castor, Marco Couto, Rui Pereira,
  Sim\~{a}o~Melo de~Sousa, and Jo\~{a}o~Paulo Fernandes.
\newblock Greenhub farmer: Real-world data for android energy mining.
\newblock In {\em Proceedings of the 16th International Conference on Mining
  Software Repositories}, MSR '19, page 171–175. IEEE Press, 2019.

\bibitem{gilbertomsc}
Gilberto Melfe.
\newblock Energy consumption of functional programs in the context of lazy
  evaluation.
\newblock Master's thesis, Universidade da Beira Interior, 2016.

\bibitem{melfe2018sblp}
Gilberto Melfe, Alcides Fonseca, and Jo\~{a}o~Paulo Fernandes.
\newblock Evaluation of the impact on energy consumption of lazy versus strict
  evaluation of haskell data-structures.
\newblock In {\em Proceedings of the XXII Brazilian Symposium on Programming
  Languages}, SBLP '18, page 83–89, New York, NY, USA, 2018. Association for
  Computing Machinery.

\bibitem{melfe2018helping}
Gilberto Melfe, Alcides Fonseca, and Jo{\~a}o~Paulo Fernandes.
\newblock Helping developers write energy efficient haskell through a
  data-structure evaluation.
\newblock In {\em 2018 IEEE/ACM 6th International Workshop on Green And
  Sustainable Software (GREENS)}, pages 9--15. IEEE, 2018.

\bibitem{rustNews}
Rust Newsletter.
\newblock {This Week in Rust 200}.
\newblock
  \url{https://this-week-in-rust.org/blog/2017/09/19/this-week-in-rust-200/}.
\newblock [Online; accessed 28-January-2021].

\bibitem{book21_jpf}
Wellington Oliveira, Hugo Matalonga, Gustavo Pinto, Fernando Castor, and
  Jo{\~a}o~Paulo Fernandes.
\newblock Small changes, big impacts: Leveraging diversity to improve energy
  efficiency.
\newblock In Mario~Piattini Coral~Calero, Marian~Moraga, editor, {\em Software
  Sustainability}. Springer, (to appear).

\bibitem{Pereira2017}
R.~Pereira, T.~Carção, M.~Couto, J.~Cunha, J.~P. Fernandes, and J.~Saraiva.
\newblock Helping {P}rogrammers {I}mprove the {E}nergy {E}fficiency of {S}ource
  {C}ode.
\newblock In {\em Proc. of the 39th International Conference on Soft. Eng.
  Companion}, ICSE-C 2017, pages 238--240. ACM, 2017.

\bibitem{Pereira2016}
R.~Pereira, M.~Couto, J.~Cunha, J.~P. Fernandes, and J.~Saraiva.
\newblock The {I}nfluence of the {J}ava {C}ollection {F}ramework on {O}verall
  {E}nergy {C}onsumption.
\newblock In {\em Proc. of 5th Int. Workshop on Green and Sustainable
  Software}, GREENS '16, pages 15--21. ACM, 2016.

\bibitem{pereira2017locating}
Rui Pereira.
\newblock Locating energy hotspots in source code.
\newblock In {\em 2017 IEEE/ACM 39th International Conference on Software
  Engineering Companion (ICSE-C)}, pages 88--90. IEEE, 2017.
\newblock ACM-SRC - Silver medal.

\bibitem{pereiraPHD}
Rui Pereira.
\newblock {\em Energyware Engineering: Techniques and Tools for Green Software
  Development}.
\newblock PhD thesis, Universidade do Minho, 2018.

\bibitem{Pereira20}
Rui Pereira, Tiago Car{\c{c}}{\~{a}}o, Marco Couto, J{\'{a}}come Cunha,
  Jo{\~{a}}o~Paulo Fernandes, and Jo{\~{a}}o Saraiva.
\newblock Spelling out energy leaks: Aiding developers locate energy
  inefficient code.
\newblock {\em Journal of Systems and Software}, 161, 2020.

\bibitem{Pereira2017SLE}
Rui Pereira, Marco Couto, Francisco Ribeiro, Rui Rua, J\'{a}come Cunha,
  Jo{\~a}o~Paulo Fernandes, and Jo{\~a}o Saraiva.
\newblock Energy {E}fficiency {A}cross {P}rogramming {L}anguages: {H}ow {D}o
  {E}nergy, {T}ime, and {M}emory {R}elate?
\newblock In {\em Proceedings of the 10th ACM SIGPLAN International Conference
  on Software Language Engineering}, SLE 2017, pages 256--267. ACM, 2017.

\bibitem{scp2021}
Rui Pereira, Marco Couto, Francisco Ribeiro, Rui Rua, J{\'a}come Cunha,
  Jo{\~a}o~Paulo Fernandes, and Jo{\~a}o Saraiva.
\newblock Ranking programming languages by energy efficiency.
\newblock {\em Science of Computer Programming}, In Press.

\bibitem{sqamia18}
Rui Pereira, Marco Couto, Francisco Ribeiro, Rui Rua, and Jo{\~a}o Saraiva.
\newblock Energyware analysis.
\newblock In {\em Proceedings of the Seventh Workshop on Software Quality
  Analysis, Monitoring, Improvement, and Applications, {SQAMIA} 2018, Novi Sad,
  Serbia, August 27-30, 2018}, 2018.

\bibitem{emse21}
Rui Pereira, Hugo Matalonga, Marco Couto, Fernando Castor, Bruno Cabral, Pedro
  Carvalho, Sim{\~a}o~Melo de~Sousa, and Jo{\~a}o~Paulo Fernandes.
\newblock {GreenHub: A Large-Scale Collaborative Dataset to Battery Consumption
  Analysis of Android Devices}.
\newblock {\em Empirical Software Engineering Journal}, 2021.
\newblock to appear.

\bibitem{Pereira2018}
Rui Pereira, Pedro Sim\~{a}o, J\'{a}come Cunha, and Jo\~{a}o Saraiva.
\newblock j{S}tanley: {P}lacing a {G}reen {T}humb on {J}ava {C}ollections.
\newblock In {\em Proceedings of the 33rd ACM/IEEE International Conference on
  Automated Software Engineering}, ASE 2018, pages 856--859. ACM, 2018.

\bibitem{alexandrePHD}
Alexandre Perez.
\newblock {\em Spectrum-based Diagnosis: Measurements, Improvements and
  Applications}.
\newblock PhD thesis, Faculdade de Engenharia, Universidade do Porto, 2018.

\bibitem{adrianomsc}
Adriano Pinto.
\newblock Memoization para poupar energia em aplicações android.
\newblock Master's thesis, Universidade NOVA de Lisboa, 2018.

\bibitem{franciscoPHD}
Francisco Ribeiro.
\newblock {\em Explaining Software Faults in Source Code}.
\newblock PhD thesis, Universidade do Minho, ongoing: started 2018.

\bibitem{ruamsc}
Rui Rua.
\newblock Greensource - repository tailored for green software analysis.
\newblock Master's thesis, Universidade do Minho, 2018.

\bibitem{ruaPHD}
Rui Rua.
\newblock {\em Green Software in the Large: Energy-driven Techniques, Tools and
  Repositories}.
\newblock PhD thesis, Universidade do Minho, ongoing: started 2018.

\bibitem{cibse19}
Rui Rua, Marco Couto, Adriano Pinto, J{\'{a}}come Cunha, and Jo{\~{a}}o
  Saraiva.
\newblock Towards using memoization for saving energy in android.
\newblock In {\em Proceedings of the {XXII} Iberoamerican Conference on
  Software Engineering, CIbSE 2019, La Habana, Cuba, April 22-26, 2019}, pages
  279--292. Curran Associates, 2019.

\bibitem{Rua2019MSR}
Rui Rua, Marco Couto, and Jo\~{a}o Saraiva.
\newblock Green{S}ource: {A} {L}arge-scale {C}ollection of {A}ndroid {C}ode,
  {T}ests and {E}nergy {M}etrics.
\newblock In {\em Proceedings of the 16th International Conference on Mining
  Software Repositories}, MSR '19, pages 176--180, Piscataway, NJ, USA, 2019.
  IEEE Press.

\bibitem{ruaMobileSoft20}
Rui Rua, Tiago Fraga, Marco Couto, and Jo\~{a}o Saraiva.
\newblock Greenspecting android virtual keyboards.
\newblock In {\em Proceedings of the IEEE/ACM 7th International Conference on
  Mobile Software Engineering and Systems}, MOBILESoft '20, page 98–108, New
  York, NY, USA, 2020. Association for Computing Machinery.

\bibitem{SantosSPK17}
M{\'{a}}rio Santos, Jo{\~{a}}o Saraiva, Zolt{\'{a}}n Porkol{\'{a}}b, and
  D{\'{a}}niel Krupp.
\newblock Energy consumption measurement of {C/C++} programs using clang
  tooling.
\newblock In Zoran Budimac, editor, {\em Proceedings of the Sixth Workshop on
  Software Quality Analysis, Monitoring, Improvement, and Applications,
  Belgrade, Serbia, September 11-13, 2017}, volume 1938 of {\em {CEUR} Workshop
  Proceedings}. CEUR-WS.org, 2017.

\bibitem{mariomsc}
Mário Santos.
\newblock Energy analysis in the codecompass system.
\newblock Master's thesis, Universidade do Minho, 2017.

\bibitem{cefp19edu}
Jo{\~a}o Saraiva.
\newblock Green software in an engineering course.
\newblock In {\em Central European Functional Programming School: 8th Summer
  School, CEFP 2013, Cluj-Napoca, Romania, June 17-21, 2019, Revised Selected
  Papers}. Springer International Publishing, (to appear).

\bibitem{impact18}
Jo{\~{a}}o Saraiva, Rui Abreu, J{\'{a}}come Cunha, and Jo{\~{a}}o~Paulo
  Fernandes.
\newblock {GreenSoftwareLab}: Towards an engineering discipline for green
  software.
\newblock {\em Impact}, 2018(1), March 2018.

\bibitem{acta2018}
Jo{\~a}o Saraiva, Marco Couto, Csaba Szabó, and Dávid Nóvak.
\newblock Towards energy-aware coding practices for android.
\newblock {\em Acta Electrotechnica et Informatica}, 18(1):19--25, 2018.

\bibitem{progCourse}
Iztok Savnik.
\newblock {Programming II}.
\newblock
  \url{https://osebje.famnit.upr.si/~savnik/predmeti/Prog2/Introduction.pdf}.
\newblock [Online; accessed 28-January-2021].

\bibitem{silvano2016antarex}
Cristina Silvano, Giovanni Agosta, Stefano Cherubin, Davide Gadioli, Gianluca
  Palermo, Andrea Bartolini, Luca Benini, Jan Martinovi{\v{c}}, Martin
  Palkovi{\v{c}}, Kate{\v{r}}ina Slaninov{\'a}, et~al.
\newblock The antarex approach to autotuning and adaptivity for energy
  efficient hpc systems.
\newblock In {\em Proceedings of the ACM International Conference on Computing
  Frontiers}, pages 288--293, 2016.

\bibitem{newstack}
The~News Stack.
\newblock {Which Programming Languages Use the Least Electricity?}
\newblock
  \url{https://thenewstack.io/which-programming-languages-use-the-least-electricity/}.
\newblock [Online; accessed 28-January-2021].

\bibitem{hackernews}
Hackernews Thread.
\newblock {Energy Efficiency across Programming Languages}.
\newblock \url{https://news.ycombinator.com/item?id=15249289}.
\newblock [Online; accessed 28-January-2021].

\end{thebibliography}

\end{document}